\documentclass[prl,nofootinbib,floats,aps,superscriptaddress,twocolumn]{revtex4}
\usepackage[hypertex]{hyperref}
\usepackage{graphicx} % Include figure files
\usepackage{amsbsy}   % Include \boldsymbol command
\usepackage{xspace}

\newcommand{\spsi}{S_{\psi K}}
\newcommand{\spsph}{S_{\psi \phi}}

\newcommand{\CP}{\ensuremath{CP}\xspace}
\def\rhobar{\bar\rho}
\def\etabar{\bar\eta}

\newcommand{\epsK}{\ensuremath{\varepsilon_K}\xspace}
\newcommand{\dms}{\ensuremath{\Delta m_s}\xspace}
\newcommand{\dmd}{\ensuremath{\Delta m_d}\xspace}

\newcommand{\beq}{\begin{equation}}
\newcommand{\eeq}{\end{equation}}
\newcommand{\beqa}{\begin{eqnarray}}
\newcommand{\eeqa}{\end{eqnarray}}
\newcommand{\nn}{\nonumber}
\begin{document}

\pagestyle{plain}  % Page numbers bottom-center

%\vspace*{.5cm}
%\preprint{ \vbox{\vspace*{.75cm} \hbox{LBNL-59996} \hbox{UCB-PTH-06/05} 
%\hbox{MIT-CTP 3734} \hbox{hep-ph/0604112} }}

\title{\boldmath Implications of the measurement of the $B^0_s\bar B^0_s$ mass
difference}

\author{Zoltan Ligeti}
\affiliation{Ernest Orlando Lawrence Berkeley National Laboratory,
University of California, Berkeley, CA 94720}
\affiliation{Center for Theoretical Physics, Massachusetts Institute of
Technology, Cambridge, MA 02139}

\author{Michele Papucci}
\affiliation{Ernest Orlando Lawrence Berkeley National Laboratory,
University of California, Berkeley, CA 94720}
\affiliation{Department of Physics, University of California,
Berkeley, CA 94720}

\author{Gilad Perez}
\affiliation{Ernest Orlando Lawrence Berkeley National Laboratory,
University of California, Berkeley, CA 94720}

%\date{\today}

\begin{abstract}                                                                                 

We analyze the significant new model independent constraints on extensions of
the standard model (SM) that follow from the recent measurements of the
$B^0_s\bar B^0_s$ mass difference.  The time-dependent \CP asymmetry in $B_s\to
\psi\phi$, $S_{\psi\phi}$, will be measured with good precision in the first
year of LHC data taking, which will further constrain the parameter space of
many extensions of the SM, in particular, next-to-minimal flavor violation.  The
\CP asymmetry in semileptonic $B_s$ decay, $A_{\rm SL}^s$, is also important to
constrain these frameworks, and could give further clues to our understanding
the flavor sector in the LHC era. We point out a strong correlation between
$S_{\psi\phi}$ and $A_{\rm SL}^s$ in a very broad class of new physics models.

\end{abstract}
%%\pacs{}
\maketitle

%%%%%%%%%%%%%%%%%%%%%%%%%%%%%%%%%%%%%%%%%%%%%%%%%%%%%%%%%%%%%%%%%%%%%%%%%%%%

%%\section{Introduction}

Recently the D\O~\cite{D0} and CDF~\cite{CDF} collaborations reported
measurements of the $B^0_s\bar B^0_s$ mass difference
\beqa\label{Bsmix}
&& 17\,{\rm ps}^{-1} < \dms < 21\,{\rm ps}^{-1} 
  ~~~\mbox{(90\%\,CL,~ D\O)}\,, \nonumber\\
&& \dms = (17.31 ^{+0.33}_{-0.18} \pm 0.07)\, {\rm ps}^{-1} ~~~\mbox{(CDF)}\,.
\eeqa
The probability of the signal being a background fluctuation is 0.2\% (5\%) for
CDF (D\O).  More important than the (moderate) improvements in the standard
model (SM) global fit for the Cabibbo-Kobayashi-Maskawa (CKM) elements is that
these measurements provide the first direct constraint on new physics (NP)
contributions to the $B_s\bar B_s$ mixing amplitude.

We focus below on a large class of NP models with the following
features~\cite{Frame}:
(I) The $3\times3$ CKM matrix is unitary;
(II) Tree-level decays are dominated by the SM contributions.
These assumptions are rather mild and allow for large
deviations from the SM predictions. It is therefore  important to examine
how present and near future experimental data constrain the
parameter space of such models.

We expect NP contributions to modify the predictions for observables that are
related to flavor-changing neutral current (FCNC)
processes. A priori, we have no knowledge of the expected size of these
contributions.  However, due to the hierarchy problem, new degrees of freedom
should be present near the electroweak symmetry breaking (EWSB) scale.  Allowing
for 10\% fine tuning in the SM Higgs potential, the new degrees of freedom which
regularize the Higgs quadratic divergence from the top-loop should have masses
$m_X \sim  3$\,TeV~(see, e.g.,~\cite{BH}). In a most generic natural theory such
a particle can have tree-level non-universal couplings to the SM quarks. Thus,
after integrating out $X$, four-fermion operators of the form $(\bar d^i d^j)^2
/ m_X^2$ ($i,j=1..3$) are expected to be generated with order one complex
coefficients. This would contribute to many well-measured processes in the
$B^0_q$ ($q=d,s$) and $K^0$ systems. For instance, in $K^0 \bar K^0$ and
$B_q^0 \bar B_q^0$ mixing we can parameterize the ratio between the NP and the
SM short distance contributions by $h_{K,q}\, e^{2i\sigma_{K,q}}$. Assuming
arbitrary \CP violating phases, we expect the following orders of magnitudes for
these parameters in the general case
\beq\label{hgen}
h_{K,d,s}^{\rm gen}\sim \left({4\pi v\over m_X\lambda^{5,3,2}}\right)^2 
  \sim {\cal O} \big( 10^5,\, 10^3,\, 10^2 \big) \,,
\eeq
where $v$ is the EWSB scale.
Clearly, such huge values are excluded by many other observables, but this way
of presenting the NP expectation will be useful in the following discussion. 
The smaller the ratio between the experimental bounds on $h_{K,d,s}$ and
$h_{K,d,s}^{\rm gen}$, the more disfavored this framework is.

The bounds on the above parameters prior to the $\dms$ measurement were given
in~\cite{NMFV}, $h_{K,d} \lesssim 0.6,\, 0.4$, which are ${\cal O} (10^{-6},\,
10^{-4})$ times smaller than Eq.~(\ref{hgen}), while no significant bound was
found on $h_s$.  The smallness of these ratios demonstrates that
generic models which address the SM fine tuning problem are in great tension
with indirect bounds from FCNC processes. These require that the scale of $m_X$
should be orders of magnitude above the TeV scale.

The SM quark flavor sector is far from being generic as well.  Most of the
SM flavor parameters are small and hierarchical, and the flavor sector possesses
an approximate $U(3)_d\times U(2)_u\times U(2)_Q$ flavor symmetry (here $d,\, u,\, Q$ correspond to the
down and up type singlet and doublet quarks, respectively).
Roughly speaking, only the top Yukawa
coupling violates these approximate symmetries. 
Thus it is not inconceivable that the NP at $m_X$ will share the same
flavor symmetries. In this case its contributions to FCNC processes
will be suppressed and Eq.~(\ref{hgen}) overestimates
their size. Below we
therefore assume that this is the case, and the new non-flavor-universal higher
dimensional operators are
invariant under these symmetries.

The special case in which these new operators are fully aligned with the SM
Yukawa matrices corresponds to the minimal flavor violation (MFV) framework.
Then the only sources of flavor and \CP violation are due to the SM~\cite{MFV}. 
A more general case is when the new operators are only quasi-aligned with the SM
Yukawa matrices, that is, in the basis where the new operators are flavor
diagonal, the diagonalizing matrices of the Yukawa couplings are at least as
hierarchical as the CKM matrix. This constitutes  next-to-minimal minimal flavor
violation (NMFV)~\cite{NMFV}.  In this case there are new flavor and \CP
violating parameters, so NMFV is almost as generic as the class of models
defined above by conditions (I) and (II). However, our assumption of
quasi-alignment provides a useful way for ``power counting'' and to estimate the
size of the expected NP contributions. Moreover it is also realized by many
supersymmetric and non-supersymmetric models (see~\cite{NMFV} for more
details), providing a powerful framework for model independent
analysis.

What is the expected size of the NP contributions?  Four-fermion
operators  are generated when the NP is integrated out at a scale of
order $\Lambda_{\rm NMFV} \sim m_X \sim 3\,$TeV.  Consider, for
example,  the operator  
$\left(\bar Q_3 Q_3/\Lambda_{\rm NMFV}\right)^2$ defined in the interaction
basis (gauge, Lorentz indices and
${\cal O}(1)$ coefficients are omitted). In the mass basis, this
operator contributes to $\Delta F=2$ processes as
%\beq
%\left[{(D_L^*)_{3i} (D_L)_{3j}\, \bar Q_i Q_j\over \Lambda_{\rm NMFV}}\right]^2
%  \sim \left[{(V_{\rm CKM}^*)_{3i} (V_{\rm CKM})_{3j}\, \bar Q_i Q_j
%  \over \Lambda_{\rm NMFV}} \right]^2.
                                %\eeq
$[{(D_L^*)_{3i} (D_L)_{3j}\, \bar Q_i Q_j/ \Lambda_{\rm NMFV}}]^2
  \sim [{(V_{\rm CKM}^*)_{3i} (V_{\rm CKM})_{3j}\, \bar Q_i Q_j
  / \Lambda_{\rm NMFV}} ]^2$, where $D_L$ is the 
  rotation matrix of the down type doublet quarks.
Comparing the NP contributions to the SM ones we find that within the NMFV
we expect
\beq\label{hNMFV}
  h_{K,d,s}^{\rm NMFV}\sim {\cal O}(1)\,.
\eeq

The magnitudes of $h_{K,d,s}$ are inversely proportional to the cutoff of the
theory  and provide a measure of the tuning in the model. Moreover, a connection
between $\Lambda_{\rm NMFV}$ and $m_X$ relates this fine tuning to the one in
the Higgs sector.  Consequently, just as in the case of electroweak precision
tests, any model of this class will be disfavored if the constraints on the
$h_{K,d,s}$ drop below the 0.1 level.

Below we focus on NP in $\Delta F=2$ processes, which are in
general theoretically cleaner and have simpler operator
structures.  To constrain deviations from the SM in these processes, the
tree-level observables $|V_{ub}/V_{cb}|$ and $\gamma$ extracted from the \CP
asymmetry in $B^\pm\to DK^\pm$ modes are crucial, because they are unaffected by
NP.   We consider in addition the following observables: the $B_q^0\bar
B_q^0$ ($q=d,s$) mass differences, $\Delta m_q$; \CP violation in $B_q^0$
mixing, $A_{\rm SL}^q$~\cite{LLNP}; the time dependent \CP asymmetries in
$B_d^0$ decays, $S_{\psi K}$ and $S_{\rho\rho,\pi\pi,\rho\pi}$; and  the
time dependent \CP asymmetry in $B_s^0$ decay, $S_{\psi\phi}$\footnote{By
$S_{\psi\phi}$ we mean the \CP asymmetry divided by $(1-2f_{\psi\phi}^{\rm
odd})$ to correct for the \CP-odd $\psi\phi$ fraction, which also equals
$-S_{\psi\eta^{(\prime)}}$.}; the lifetime difference between the $CP$-even
and $CP$-odd $B_s$ states, $\Delta\Gamma_s^{CP}$~\cite{Grossman:1996er}.  (Of
these, $A_{\rm SL}^s$ and $S_{\psi\phi}$ have not been measured, however, they
will be important in the discussion below.)  

The NP contributions to $B_d^0$ and $B_s^0$ mixing can be expressed in terms of
four parameters, $h_q$ and $\sigma_q$ defined by $M_{12}^q = (1+h_q
e^{2i\sigma_q}) M_{12}^{q,{\rm SM}}$, where $M_{12}^{q,{\rm SM}}$ is the
dispersive part of the $B^0_q\bar B^0_q$ mixing amplitude in the SM.  (For a
similar parameterization of NP in the $K^0$ system, see~\cite{NMFV}.)  Then the
predictions for the above observables are modified compared to the SM as
follows:
\beqa\label{par}
\Delta m_q &=& \Delta m_q^{\rm SM}\,
  \big|1+h_q e^{2i\sigma_q}\big| , \nn\\
\spsi &=& \sin \big[2\beta + \arg\big(1+h_d e^{2i\sigma_d}\big)\big]\,, \nn\\
\spsph &=& \sin \big[ 2\beta_s - \arg \big(1+h_s
    e^{2i\sigma_s}\big)\big]\,,\nn\\
A^q_{\rm SL} &=& {\rm Im}\, \big\{ \Gamma_{12}^q / \big[ M_{12}^{q,{\rm SM}}
  (1+h_q e^{2i\sigma_q}) \big] \big\} ,\nn\\
\Delta\Gamma_s^{CP} &=& \Delta\Gamma_s^{\rm SM}\,
  \cos^2 \big[\arg\big(1+h_s e^{2i\sigma_s}\big)\big] .
\eeqa
Here $\lambda \approx 0.23$ is the Wolfenstein parameter, $\beta_s =
\arg[-(V_{ts}V_{tb}^*)/(V_{cs}V_{cb}^*)] \approx 1^\circ$ is the angle of a
squashed unitarity triangle, and $\Gamma_{12}^q$ is the absorptive part of the
$B^0_q\bar B^0_q$ mixing amplitude, which is probably not significantly
affected by NP.  (We neglect ${\cal O}\big(M_W^2/\Lambda^2_{\rm NMFV}\big)$
corrections due to NP contributions to SM tree-level $\Delta F=1$ processes; for
a different approach, see~\cite{Bona:2005eu}.)

Looking at Eq.~(\ref{par}) one notices a fundamental
difference between the $B_d$ and $B_s$ systems. 
The SM contributions affecting the $B_d$ system are
related to the non-degenerate unitarity triangle.
Thus the determination of $h_d,\sigma_d$ is strongly correlated
with that of the Wolfenstein parameters, $\bar\rho,\bar\eta$.
On the other hand the unitarity triangle relevant for the $B_s$ system
is nearly degenerate and therefore the determination of $h_s,\sigma_s$
is almost independent of $\bar \rho,\bar \eta$.

\begin{figure*}[t]
\centerline{\includegraphics[width=.95\columnwidth]{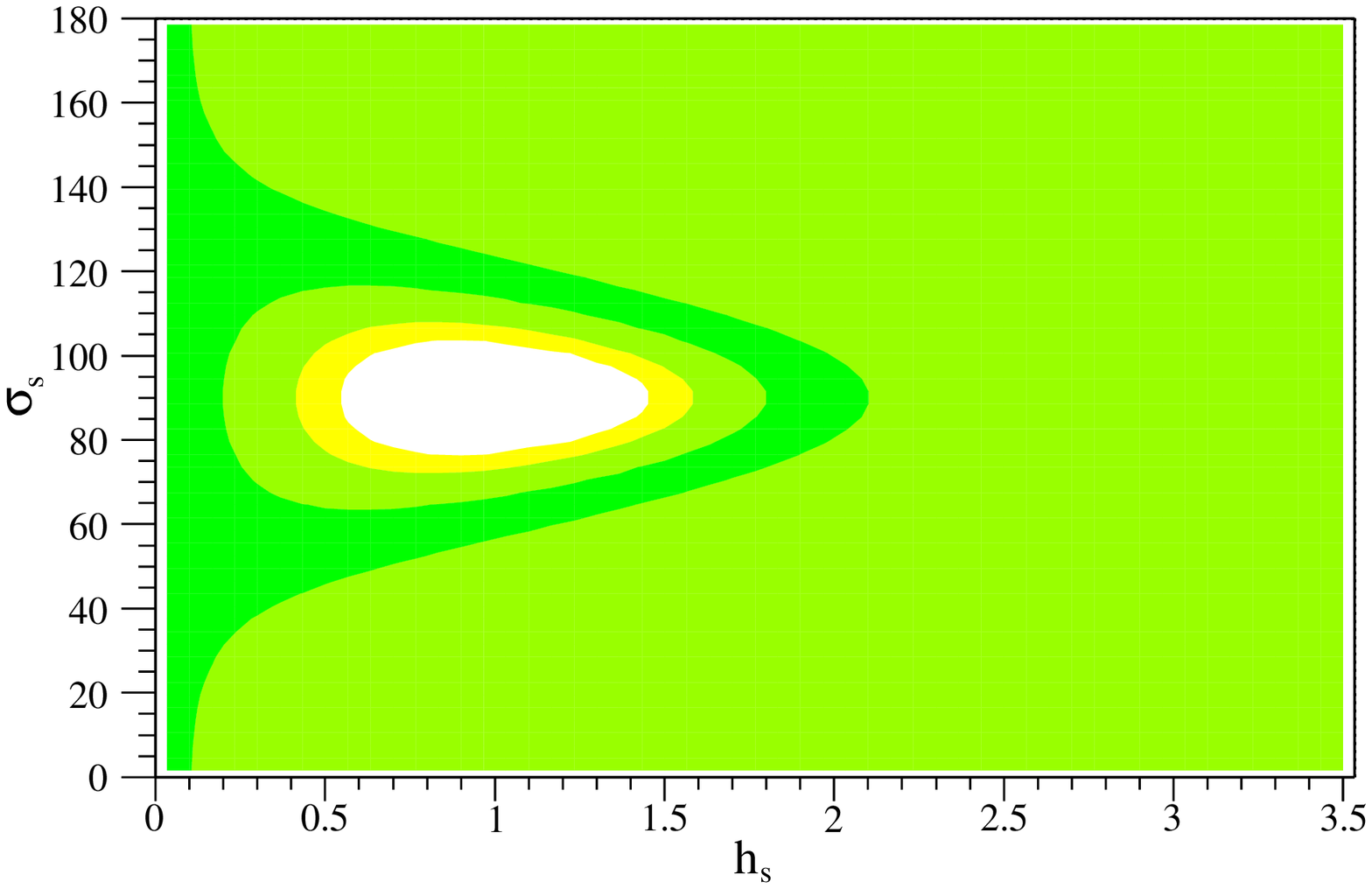} \hfil
  \includegraphics[width=.95\columnwidth]{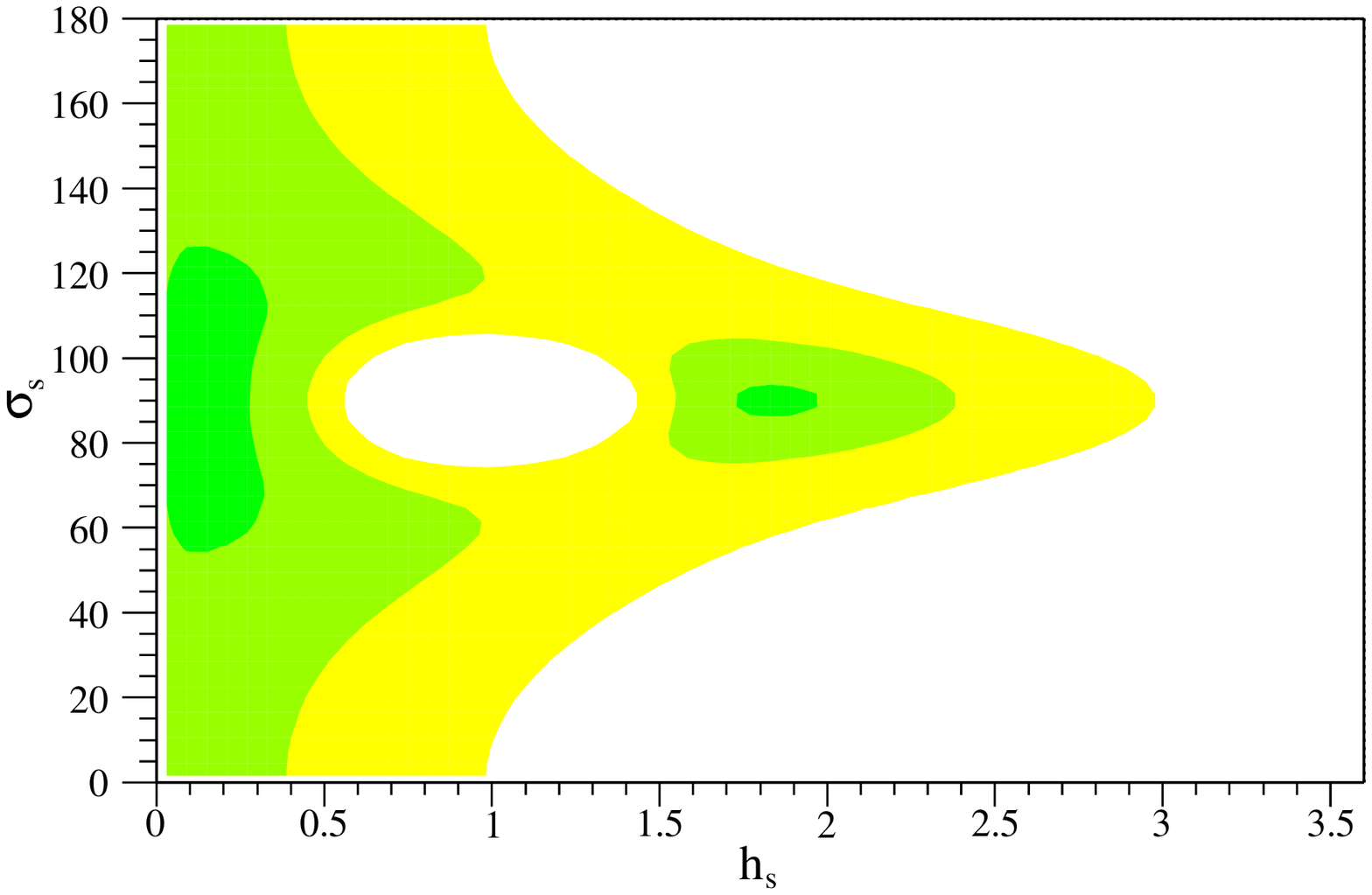}}
\caption{The allowed range for $h_s,\sigma_s$ using the data before (left) and
after (right) the recent $\dms$ and $\Delta\Gamma_s$ measurements.  For $\dms$
only the CDF result was used.  The dark, medium, and light shaded areas have CL
$>$ 0.90, 0.32, and 0.05, respectively.}
\label{fighsss}
\end{figure*}

Figure~\ref{fighsss} shows the allowed $h_s,\sigma_s$ parameter space without
(left) and with (right) the measurement of $\dms$ in Eq.~(\ref{Bsmix}) and the
bound on $\Delta\Gamma_s^{CP}$, using the CKMfitter
package~\cite{ckmfitter}.\footnote{Unless otherwise stated, the input parameters
are as in~\cite{ckmfitter}.}  We used the constraint on the ratio
\beq
{\dmd \over \dms} = 
  \bigg|{1+h_d e^{2i\sigma_d} \over 1+h_s e^{2i\sigma_s}}\bigg|\, 
  \bigg|{V_{td} \over V_{ts}}\bigg|^2\,
  {m_{B_d} \over m_{B_s}}\, \xi^2\,,
\eeq
which is theoretically cleaner than either $\dmd$ or $\dms$.  Since $\dmd$
depends on $h_d,\, \sigma_d,\, \rhobar,\, \etabar$, in order to produce the
above plots these parameters were scanned over. We can easily see that the new
measurement excludes a large part of the previously allowed parameter space. 
The excluded region around $h_s=1$ and $\sigma_s = 90^\circ$ would give
cancelling contributions to $\Delta m_s$.  The decrease in CL around $h_s=1$ is
due to the $\Delta\Gamma_s^{CP}$ constraint, which is useful at present, largely
because its central value disfavors any deviation from the SM. After a year of
LHC data, the bound from this quantity will probably be less important, because
of theoretical uncertainties.

%We use $A_{\rm SL}^d = -(0.4 \pm 5.5) \times 10^{-3}$~\cite{Grossman:2006ce}

The magnitudes of the $h_i$'s provide a measure of how much fine tuning is
required to satisfy the experimental constraints.  Generically we do not expect
the NP contributions to be MFV-like, i.e., aligned with the SM. Thus we are
interested in finding the allowed ranges of $h_i$, for $\sigma_i$ not near 0 mod
$\pi/2$. The present constraints are roughly
\beq\label{his}
h_d \lesssim 0.3\,, \qquad h_s \lesssim 2\,, \qquad h_K \lesssim 0.6\,.
\eeq

%%\section{Future prospects} 

Let us now discuss some implications of the above results. Equation~(\ref{his})
shows that at present none of the bounds on the NP parameters have reached the
0.1 level, so NMFV survives the current tests.  It is then interesting to ask
which future measurements will be most important to verify or disfavor the NMFV
framework.  The constraints on $h_{d,K}$, even though they underwent 
significant improvements in the last few years due to new SM tree-level
measurements~\cite{Ligeti04}, are now limited by the statistical errors in the
measurements of $\gamma$ (and effectively $\alpha$) and the hadronic parameters
in the determination of $|V_{ub}|$ from semileptonic decays and $|V_{td}|$ from
$\dmd$. The improvements in these constraints will be incremental, as they
depend on the integrated luminosities at the $B$ factories and on progress in
lattice QCD. The constraint from $\epsK$ on the $K$ system is also dominated by
hadronic uncertainties. At present, the bound on $h_s$ is weaker than that on
$h_d$, since only one measurement, $\dms$, constrains it, and the hadronic
uncertainties are comparable.  

\begin{figure}[t]
\includegraphics[width=0.95\columnwidth]{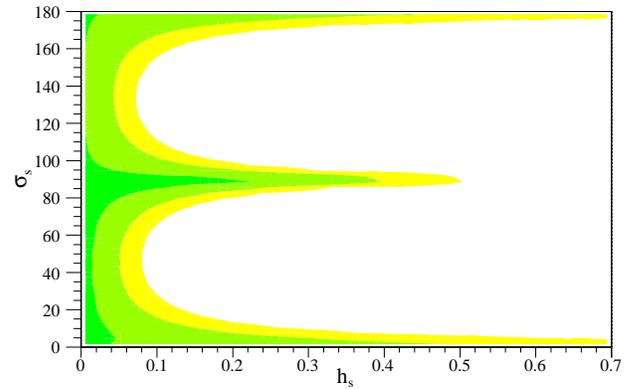}
\caption{The allowed range for $h_s,\sigma_s$ using the 1 year LHCb projection,
assuming the SM prediction as the central value.}
\label{fighsss_future}
\end{figure}

However the $B_s$ system is exceptional because a measurement of
$S_{\psi\phi}$ (or a strong bound on it) would provide a very sensitive test of
NMFV, which is neither obscured by hadronic uncertainties nor by uncertainties
in the CKM parameters.  In the SM $S_{\psi\phi}$ is suppressed by $\lambda^2$
(the SM CKM fit gives $\sin2\beta_s = 0.038 \pm 0.003$), whereas
Eq.~(\ref{par}) implies
\beq
S_{\psi\phi} = - {h_s\sin(2\sigma_s) \over | 1 + h_s e^{2i\sigma_s}|}
  + \sin(2\beta_s) {1 + h_s \cos(2\sigma_s) \over | 1 + h_s e^{2i\sigma_s}|} \,,
\eeq
where we set $\cos2\beta_s$ to unity. Thus when the sensitivity of the
measurement of $S_{\psi\phi}$ reaches the SM level, it will provide us with a
strong test of NMFV. The precision that will be achieved in forthcoming
experiments depends on the value of $\dms$, but since we now know $\dms$, we can
use the LHC projections for the SM case.  LHCb expects to reach $\sigma
(S_{\psi\phi}) \approx 0.03$ with the first year (2\,fb$^{-1}$) data~\cite{LHCb}
(in several years the uncertainty may be reduced to 0.01). 
Figure~\ref{fighsss_future} shows the resulting constraint on $h_s,\sigma_s$,
assuming an experimental measurement $S_{\psi\phi} = 0.04 \pm 0.03$. This plot
demonstrates that already with one year of data the bound on $h_s$ will be
better than 0.1, which will severely constrain the NMFV class of models.  Even
initial data on $S_{\psi\phi}$ at the Tevatron may constrain new physics in
$B_s$ mixing comparable to similar bounds on $h_d,\sigma_d$ in the $B_d$ sector.

\begin{figure}[t]
\includegraphics[width=.95\columnwidth]{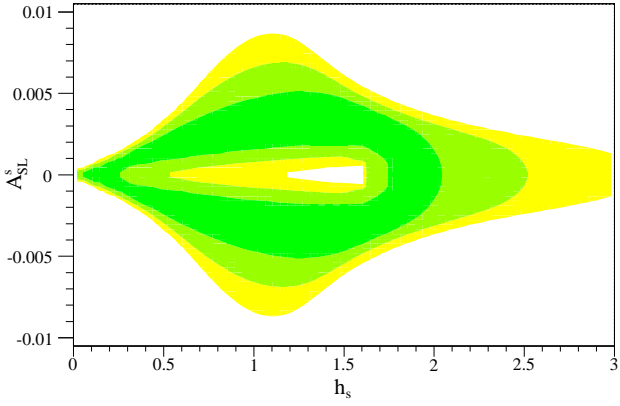}   
\caption{The current allowed range of $A_{\rm SL}^s$ as a function of $h_s$.}
\label{fighsasls}
\end{figure}

Another sensitive probe of this class of models is the \CP asymmetry in
semileptonic $B_s$ decays, $A_{\rm SL}^s$. In the SM it is unobservably small,
because the short distance contributions are much larger than the long distance
part, $|\Gamma^s_{12}/M^s_{12}| \propto m_b^2/m_t^2$, and the two
contributions are highly aligned, $\arg(\Gamma^s_{12}/M^s_{12}) \propto (m_c^2 /
m_b^2) \sin 2\beta_s$~\cite{LLNP}.  Given the new $\dms$ result, we know that
even in the presence of NP the first suppression factor can only be moderately
affected, while the second one can be significantly enhanced in the presence of
new \CP violating phases.  Figure~\ref{fighsasls} shows the allowed range of
$A_{\rm SL}^s$, taking into account the new constraint from $\dms$.  We find
\beq
 A_{\rm SL}^s < 0.01\,,
\eeq
which extends three order of magnitude above the SM
prediction~\cite{Beneke:2003az}. In particular, $|A_{\rm SL}^s| > |A_{\rm
SL}^d|$ is possible, contrary to the SM, in which $|A_{\rm SL}^s / A_{\rm SL}^d|
\approx |V_{td}/V_{ts}|^2$.  This demonstrates that while the constraint from
the $\dms$ measurement is of great importance, it still leaves plenty of room
for deviations from the SM within NMFV.

\begin{figure}[t]
\includegraphics[width=.95\columnwidth]{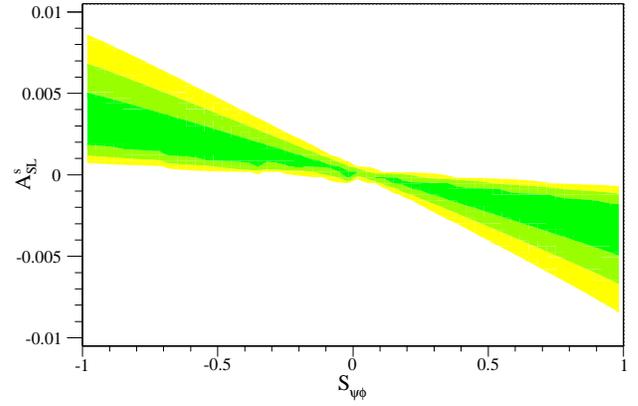}   
\caption{The correlation between $A_{\rm SL}^s$ and $S_{\psi\phi}$.}
\label{fighsasls_lhc}
\end{figure}

Finally we point out that $A_{\rm SL}^s$ and $S_{\psi\phi}$  are highly
correlated in the region in which $h_s,\sigma_s\gg \beta_s$ and $h_s$ is
moderate.  Defining $2\theta_s \equiv \arg(1+h_s e^{2i\sigma_s})$, we
have $S_{\psi\phi} = \sin(2\beta_s-2\theta_s)$, so $A_{\rm SL}^s$
can be written as 
\beq\label{a1}
A_{\rm SL}^s = \bigg|{\Gamma^s_{12}\over M^s_{12}} \bigg|^{\rm SM} 
  \sin(2\theta_s) + {\cal O}\bigg(h_s^2, {m_c^2\over m_b^2}\bigg)\,.
\eeq
Thus, we find
\beq 
A_{\rm SL}^s = - \left|\Gamma^s_{12}\over M^s_{12}\right|^{\rm SM} 
  S_{\psi\phi} + {\cal O}\bigg(h_s^2, {m_c^2\over m_b^2}\bigg)\,.
\eeq
Figure~\ref{fighsasls_lhc} shows $A_{\rm SL}^s$ as a function of $S_{\psi\phi}$,
taking into account the constraint from $\dms$ [without neglecting the ${\cal
O}(h_s^2, m_c^2 / m_b^2)$ terms].  As explained above, the two observables are
strongly correlated. Deviation from this prediction would provide a clear
indication of new physics beyond the generic framework defined by (I) and~(II).

%%%%%%%%%%%%%%%%%%%%%%%%%%%%%%%%%%%%%%%%%%%%%%%%%%%%%%%%%%%%%%%%%%%%%%%%%%%%

Acknowledgments: Sign mistakes in v1 of this paper were first pointed out in v2 of~\cite{Burasms}.
%Just before replacing v1 of this paper with v2, v2 of~\cite{Burasms}
%pointed out sign mistakes in our v1, which our v2 agreed with.
%When v1 of this paper was about to be replaced with v2, v2 of~\cite{Burasms}
%pointed out sign mistakes in our v1, which our v2 agreed with.
%% in agreement with our v2.
We are grateful to Daria Zieminska for drawing our attention to the relevance 
of $\Delta\Gamma_s^{CP}$. While we were including it in our analysis, 
Refs.~\cite{Grossman:2006ce,Bona:2006sa} appeard with similar results.
We thank A.~Cerri, A.~H\"ocker and C.~Paus  for helpful discussions, 
and O.~Schneider for pointing out typos in v1.
This work was supported in part by the 
Director, Office of Science, Office of High Energy Physics of the 
U.S.\ Department of Energy under contract DE-AC02-05CH11231.

%%%%%%%%%%%%%%%%%%%%%%%%%%%%%%%%%%%%%%%%%%%%%%%%%%%%%%%%%%%%%%%%%%%%%%%%%%%%


\vspace*{-.1cm}

\begin{thebibliography}{99}

%\cite{Abazov:2006dm}
\bibitem{D0}
  V.~Abazov {\it et al.} [D\O\ Collaboration],
  %``First direct two-sided bound on the B/s0 oscillation frequency,''
  hep-ex/0603029.
  %%CITATION = HEP-EX 0603029;%%

\bibitem{CDF}
  A.~Abulencia {\it et al.} [CDF Collaboration],
  %``Measurement of the Bs-Bsbar Oscillation Frequency,''
  hep-ex/0606027.
  %%CITATION = HEP-EX 0606027;%%

\bibitem{Frame}
C.~Dib, D.~London and Y.~Nir,
%``CP Asymmetries In B0 Decays Beyond The Standard Model,''
Int.\ J.\ Mod.\ Phys.\ A {\bf 6}, 1253 (1991);
%%CITATION = IMPAE,A6,1253;%%

\bibitem{BH}
  R.~Barbieri, T.~Gregoire and L.~J.~Hall,
  %``Mirror world at the Large Hadron Collider,''
  hep-ph/0509242.
  %%CITATION = HEP-PH 0509242;

\bibitem{NMFV}
  K.~Agashe, M.~Papucci, G.~Perez and D.~Pirjol,
  %``Next to minimal flavor violation,''
  hep-ph/0509117.
  %%CITATION = HEP-PH 0509117;%%

\bibitem{MFV}
 A.~Ali and D.~London,
  %``Profiles of the unitarity triangle and CP-violating phases in the  standard
  %model and supersymmetric theories,''
  Eur.\ Phys.\ J.\ C {\bf 9}, 687 (1999)
  [hep-ph/9903535];
  %%CITATION = HEP-PH 9903535;%%
  A.~J.~Buras, P.~Gambino, M.~Gorbahn, S.~Jager and L.~Silvestrini,
  %``Universal unitarity triangle and physics beyond the standard model,''
  Phys.\ Lett.\ B {\bf 500}, 161 (2001)
  [hep-ph/0007085];
  %%CITATION = HEP-PH 0007085;%%
G.~D'Ambrosio, G.~F.~Giudice, G.~Isidori and A.~Strumia,
%``Minimal flavour violation: An effective field theory approach,''
Nucl.\ Phys.\ B {\bf 645}, 155 (2002)
[hep-ph/0207036].
%%CITATION = HEP-PH 0207036;%%

\bibitem{LLNP}
  S.~Laplace, Z.~Ligeti, Y.~Nir and G.~Perez,
  %``Implications of the CP asymmetry in semileptonic B decay,''
  Phys.\ Rev.\ D {\bf 65}, 094040 (2002)
  [hep-ph/0202010].
  %%CITATION = HEP-PH 0202010;%%

\bibitem{Grossman:1996er}
  Y.~Grossman,
  %``The B/s width difference beyond the standard model,''
  Phys.\ Lett.\ B {\bf 380}, 99 (1996)
  [hep-ph/9603244];
  %%CITATION = HEP-PH 9603244;%%
%%\bibitem{Dunietz:2000cr}
  I.~Dunietz, R.~Fleischer and U.~Nierste,
  %``In pursuit of new physics with B/s decays,''
  Phys.\ Rev.\ D {\bf 63}, 114015 (2001)
  [hep-ph/0012219].
  %%CITATION = HEP-PH 0012219;%%

\bibitem{Bona:2005eu}
  M.~Bona {\it et al.}  [UTfit Collaboration],
  %``The UTfit collaboration report on the status of the unitarity triangle
  %beyond the standard model. I: Model-independent analysis and minimal flavour
  %violation,''
  JHEP {\bf 0603}, 080 (2006)
  [hep-ph/0509219].
  %%CITATION = HEP-PH 0509219;%%

\bibitem{ckmfitter}
A.~Hocker, H.~Lacker, S.~Laplace and F.~Le Diberder,
%``A new approach to a global fit of the CKM matrix,''
Eur.\ Phys.\ J.\ C {\bf 21} (2001) 225
[hep-ph/0104062];
%%CITATION = HEP-PH 0104062;%%
%%\bibitem{Charles:2004jd}
  J.~Charles {\it et al.},
  %``CP violation and the CKM matrix: Assessing the impact of the asymmetric B
  %factories,''
  Eur.\ Phys.\ J.\ C {\bf 41} (2005) 1
  [hep-ph/0406184];
  %%CITATION = HEP-PH 0406184;%%
  updates at \url{http://ckmfitter.in2p3.fr/}.

\bibitem{Ligeti04}
Z.~Ligeti,
%``The CKM matrix and CP violation,''
hep-ph/0408267. 
%%CITATION = HEP-PH 0408267;%%

\bibitem{LHCb}
O.~Schneider, Talk at the ``Flavour in the era of the LHC" workshop, Nov.\ 2005,
CERN, \url{http://indico.cern.ch/conferenceDisplay.py?confId=a052129}.

\bibitem{Beneke:2003az}
M.~Beneke, G.~Buchalla, A.~Lenz and U.~Nierste,
%``CP asymmetry in flavor specific B decays beyond leading logarithms,''
Phys.\ Lett.\ B {\bf 576} (2003) 173
[hep-ph/0307344];\\
%%CITATION = HEP-PH 0307344;%%
%\bibitem{Ciuchini:2003ww}
M.~Ciuchini, E.~Franco, V.~Lubicz, F.~Mescia and C.~Tarantino,
%``Lifetime differences and CP violation parameters of neutral B mesons at the
%next-to-leading order in QCD,''
JHEP {\bf 0308} (2003) 031
[hep-ph/0308029].
%%CITATION = HEP-PH 0308029;%%

\bibitem{Burasms}
  M.~Blanke, A.~J.~Buras, D.~Guadagnoli and C.~Tarantino,
  %``Minimal Flavour Violation Waiting for Precise Measurements of Delta Ms,
  %$|$Vub$|$, gamma and B^0_{sd} $\to$ mu+ mu-,''
  hep-ph/0604057.
  %%CITATION = HEP-PH 0604057;%%

\bibitem{Grossman:2006ce}
  Y.~Grossman, Y.~Nir and G.~Raz,
  %``Constraining the phase of B/s - anti-B/s mixing,''
  hep-ph/0605028.
  %%CITATION = HEP-PH 0605028;%%

\bibitem{Bona:2006sa}
  M.~Bona {\it et al.}  [UTfit Collaboration],
  %``The UTfit Collaboration Report on the Unitarity Triangle beyond the
  %Standard Model: Spring 2006,''
  hep-ph/0605213.
  %%CITATION = HEP-PH 0605213;%%

\end{thebibliography}
\end{document}